%% file: main.tex
\documentclass[cameraready]{Interspeech}

\title{Breaking Shortcut Learning for Cross-Trial EEG-Guided Target Speech Extraction via Two-Stage Training}

\author[affiliation={1}, orcid=0009-0001-1056-5916]{Wonchul}{Shin}
\author[affiliation={2}, orcid=0000-0002-6663-9152]{Inyong}{Choi}
\author[affiliation={1,3,4}, orcid=0000-0002-4210-0312, correspondingauthor]{Kyogu}{Lee}

\address{
    $^1$ Department of Intelligence and Information, Seoul National University, Republic of Korea \\
    $^2$ Department of Communication Sciences and Disorders, University of Iowa, United States \\
    $^3$ Interdisciplinary Program in Artificial Intelligence, Seoul National University, Republic of Korea \\
    $^4$ Artificial Intelligence Institute, Seoul National University, Republic of Korea
}

\email{swc0406@snu.ac.kr, inyong-choi@uiowa.edu, kglee@snu.ac.kr}


\usepackage{comment}
\usepackage{subcaption}
\usepackage{multirow}
\usepackage{booktabs}
\usepackage{makecell}
\usepackage{pdflscape}

\begin{document}

\maketitle

\begin{abstract}
Recent end-to-end models for EEG-guided target speech extraction report impressive results, underscoring potential for neuro-steered hearing technologies. However, our analysis reveals that high within-trial performance can be driven by trial-specific EEG structure that acts as shortcuts for target selection, leading to poor generalization on unseen trials. To overcome this gap, we propose TRUST-TSE, a two-stage framework to mitigate shortcut learning. By introducing contrastive pretraining with attended-speaker negative sampling, we encourage the EEG encoder to capture fine-grained EEG--speech alignment while suppressing trial-identity cues. We also employ a confidence-weighted extraction objective based on EEG--source similarity to guide extraction using the learned representations. Experiments on KUL and DTU datasets show that TRUST-TSE outperforms end-to-end baselines under strict cross-trial protocols, addressing a key reliability bottleneck of existing approaches.
\end{abstract}

\section{Introduction}

Understanding speech in noisy environments requires selective attention to an intended auditory stream while suppressing competing sources. This challenge, commonly referred to as the cocktail party problem, is particularly consequential for individuals with hearing loss, for whom competing sounds can severely degrade speech intelligibility and listening comfort~\cite{mcdermott2009cocktail,bronkhorst2000cocktail}. While contemporary hearing aids can improve audibility, they often fall short of providing attention-dependent selectivity in multi-source scenes because these devices have no direct access to which source the listener is attending to at a given moment. This limitation has motivated neuro-steered approaches that decode the attended source from neural activity~\cite{mesgarani2012selective,ding2012emergence}. Among candidate neural interfaces, electroencephalography (EEG) has attracted substantial research interest due to its practicality and non-invasive nature~\cite{o2015attentional,mirkovic2015decoding,geirnaert2021electroencephalography}.

\begin{figure}[t]
  \centering

  \begin{subfigure}[t]{\linewidth}
    \centering
    \includegraphics[width=\linewidth]{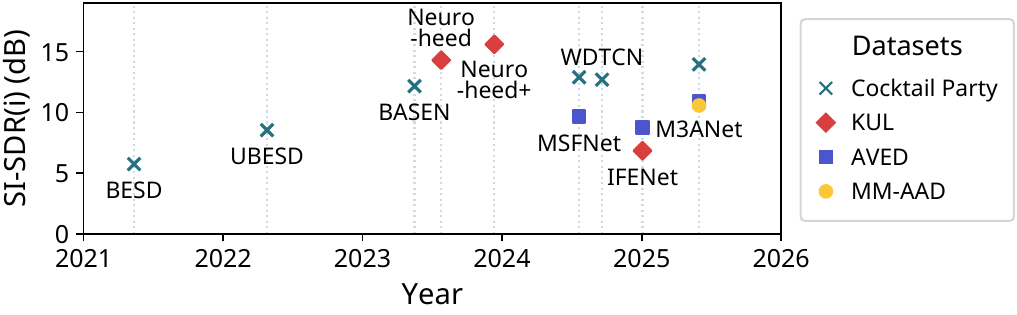}
    \caption{}
    \label{fig:fig1a}
  \end{subfigure}

  \vspace{0.5em}

  \begin{subfigure}[t]{\linewidth}
    \centering
    \includegraphics[width=\linewidth]{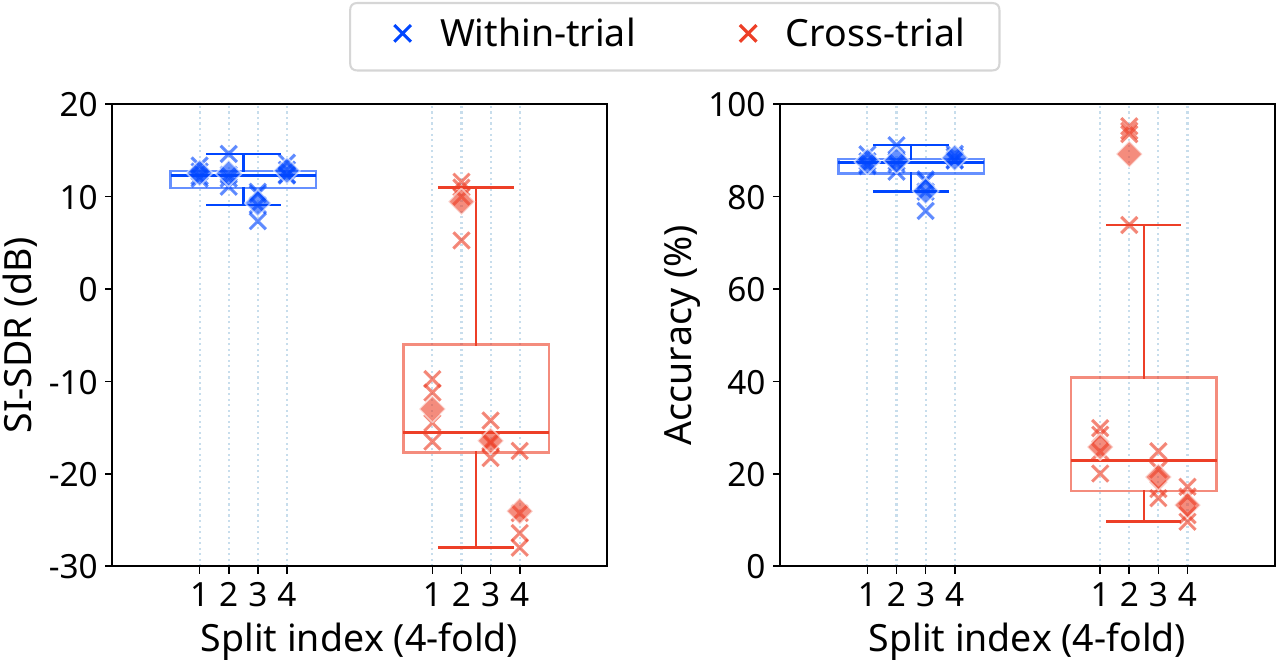}
    \caption{}
    \label{fig:fig1b}
  \end{subfigure}

  \caption{(a) Reported SI-SDR values over publication years for end-to-end EEG-guided TSE models. Values are taken from the original papers under their evaluation protocols~\cite{hosseini2021speaker,hosseini2022end,zhang2023basen,pan2024neuroheed,pan2024neuroheed+,fan2024msfnet,zuo2025geometry,fan2025improved,ijcai2025p894}. (b) Within-trial vs. cross-trial evaluation of NeuroHeed on KUL. Split indices (1–4) denote folds for each protocol. X markers show 4 seeds per fold, diamonds are fold means, and boxplots summarize all runs per protocol.}
  \label{fig:fig1}
\end{figure}

Recent studies increasingly adopt end-to-end deep learning for EEG-guided target speech extraction (TSE), in which neural networks jointly process EEG and the acoustic mixture to separate the attended speech from competing sources~\cite{hosseini2021speaker,hosseini2022end,zhang2023basen}. These end-to-end systems have shown promising results, with state-of-the-art models reporting average SI-SDR (Scale-Invariant Signal-to-Distortion Ratio) above 10 dB and attended-source accuracy close to 90\% (Figure~\ref{fig:fig1a}). These results indicate that the models can produce a clearly separated speech estimate from a multi-talker mixture, and the estimate corresponds to the attended source in most cases. However, despite these impressive numbers, their practical implications may remain limited once evaluation moves beyond the specific protocol under which they are reported. In particular, prior work across EEG decoding tasks, including spatial auditory attention decoding (AAD)~\cite{yan2025overestimated} and image decoding~\cite{xu2026impacts}, has cautioned that commonly used \emph{within-trial} evaluation setups can overstate performance. Here, a trial denotes a continuous recording block in which the relevant task target remains fixed. For example, most EEG auditory attention datasets consist of multiple trials, and in each trial participants listen to continuous speech while attending to a designated target, such as a specific speaker or a spatial direction. Importantly, continuous EEG often shows temporal autocorrelation over seconds to minutes~\cite{linkenkaer2001long,nikulin2004long}, and slow drifts in nuisance factors such as electrode impedance can introduce nonstationary trends that spuriously increase the similarity between neighboring segments~\cite{kappenman2010effects}. Consequently, under within-trial evaluation, models are prone to rely on temporally correlated, non-informative EEG structure rather than attention-related neural signatures, potentially leading to degraded performance on unseen trials~\cite{yan2025overestimated,rotaru2024we}. 

A practical EEG-guided TSE system must generalize to unseen trials, since real-world deployment involves trials that differ from those observed during training. Building on this motivation, we analyze why prior end-to-end training approaches fail to generalize under strict \emph{cross-trial} evaluation and propose TRUST-TSE (Trial-Robust Target Speech Extraction), a two-stage training framework that achieves reliable cross-trial generalization. Concretely, our analysis indicates that end-to-end models can minimize loss by exploiting easily learnable trial-specific EEG patterns. Since the attended speaker is fixed within each trial, these patterns can serve as a trial identifier, enabling a shortcut mapping from trial identity to the target speaker. To prevent such shortcut learning, TRUST-TSE separates EEG representation learning from EEG-guided speech extraction. In Stage 1, we pretrain an EEG encoder via contrastive learning that matches EEG segments to the corresponding attended speech segments. Crucially, we introduce attended-speaker negative sampling, where negatives are drawn from other segments of the same attended speaker, thereby discouraging shortcuts that rely on trial-specific EEG patterns. In Stage 2, we train an EEG-guided target speech extractor conditioned on the pretrained EEG embeddings. This stage uses a confidence-weighted SI-SDR objective based on EEG--source similarity, which encourages learning from samples where the EEG embedding shows a clear alignment with the attended source. Together, these components form a simple and effective alternative to end-to-end training for cross-trial generalization.

Our contributions are summarized as follows.
\begin{itemize}
\item We present a systematic analysis of cross-trial generalization failure in end-to-end EEG-guided TSE.
\item We introduce contrastive pretraining of an EEG encoder with attended-speaker negative sampling, where negatives are other non-aligned segments from the same attended speaker. We validate its advantage experimentally over standard negative sampling strategies.
\item We propose a confidence-weighted SI-SDR objective that explicitly couples pretrained EEG embeddings to target speech extraction, and we show that it is critical for making two-stage training effective in practice.
\end{itemize}
Overall, our findings highlight cross-trial generalization as a key bottleneck in the prevailing end-to-end paradigm for EEG-guided TSE, and point to two-stage training as a reliable approach to address it. TRUST-TSE serves as a principled starting point for future work on trial-robust EEG-guided extraction. The source code is publicly available at https://github.com/argaaw/TRUST-TSE.

\section{Related work}
Recent studies in spatial AAD, focusing on classifying the attended direction from EEG, have shown that standard within-trial evaluation can overestimate performance by allowing models to exploit temporal autocorrelation and trial-dependent artifacts instead of extracting attention-relevant features~\cite{yan2025overestimated,rotaru2024we}. Xu et al. substantiated this concern by demonstrating that decoding accuracy is inflated when temporal windows from the same labeled continuous block are distributed across both training and test sets~\cite{xu2026impacts}. Overall, these works primarily highlight evaluation protocol pitfalls, rather than proposing training procedures that ensure cross-trial robustness. Building on this line of work, we analyze trial-linked shortcut learning in EEG-guided TSE under end-to-end optimization and propose a two-stage training framework designed to break this shortcut and improve cross-trial generalization.

A prominent example of a two-stage TSE framework is Brain-Informed Speech Separation (BISS)~\cite{ceolini2020brain}, which first reconstructs an attended speech envelope from EEG and subsequently utilizes it as an auxiliary cue to guide the separation network. Subsequent work has increasingly shifted toward end-to-end paradigms that directly condition a speech extractor on EEG and optimize waveform-level separation objectives, achieving strong reported results under commonly used evaluation protocols~\cite{hosseini2021speaker,hosseini2022end,pan2024neuroheed,ijcai2025p894}. Our approach revisits the two-stage design but diverges fundamentally in objective and mechanism. Instead of relying on an explicit envelope proxy, we propose cross-modal contrastive pretraining to learn shortcut-resistant EEG representations, which are then used as frozen embeddings to robustly guide the target speech extraction.

\section{Diagnosing cross-trial generalization failure: a case study}
\label{sec:diagnosing}
In this section, we provide an in-depth case study that sets up the key empirical question of this paper: how well does an end-to-end EEG-guided TSE system generalize under strict cross-trial evaluation, and what drives its failure modes? For this investigation, we focus on NeuroHeed~\cite{pan2024neuroheed} and the KUL dataset~\cite{biesmans2016auditory,vandecappelle2021eeg}. NeuroHeed is a representative recent end-to-end EEG-guided TSE model that reports strong within-trial performance, and KUL is a widely used public dataset in AAD research. We later extend the evaluation to multiple datasets and methods in Section~\ref{experiments}, where we quantify that similar cross-trial degradation consistently emerges.

\subsection{KUL dataset and evaluation protocols}
\label{subsection:kul}
We describe the KUL dataset with an emphasis on its trial structure, which we use to define within-trial and cross-trial evaluation protocols throughout this paper. For a controlled comparison, we use exactly the same version of the KUL dataset as NeuroHeed~\cite{pan2024neuroheed}. In this dataset, EEG is recorded during multiple listening trials in which two concurrent Dutch speech streams are presented to the left and right ears, and subjects attend to one target stream while ignoring the competing stream. The stimuli comprise eight continuous speech streams, obtained by segmenting four Dutch stories narrated by three speakers into approximately 6-minute parts. Each trial presents a pair of streams from different speakers (Table~\ref{tab:kul}).

We next formalize two evaluation protocols that differ only in whether training and test segments may come from the same trial.  Following NeuroHeed, in the within-trial protocol, we partition each trial into temporally non-overlapping portions with a 6:1:1 split for train/validation/test. For a statistically more reliable estimate, we construct four mutually exclusive folds under this protocol, each using a different non-overlapping portion of the trial for validation and test. For the cross-trial protocol, we keep the same 6:1:1 train/validation/test ratio by assigning each trial to exactly one set. As shown in Table~\ref{tab:kul}, each pair of streams appears in two trials with the attended stream swapped, so a naive trial split can result in leakage of identical acoustic content across sets. To avoid such leakage, we construct four folds by holding out paired trials and forming (validation, test) pairs as $(1,5)$, $(2,6)$, $(3,7)$, and $(4,8)$. This ensures that no speech stream seen during training appears in validation or test, and that validation-to-test transfer cannot be attributed to memorizing a fixed attended target because the attended stream differs within each held-out pair.

\begin{table}[t]
  \caption{Stimulus pairings for the KUL dataset trials. All 16 subjects followed the same pairing and trial order, and the attended ear was counterbalanced across subjects~\cite{vandecappelle2021eeg}.}
  \label{tab:kul}
  \centering
  \resizebox{0.9\linewidth}{!}{
  \begin{tabular}{rll}
    \toprule
     \textbf{Trial} & \textbf{Attended stimulus} & \textbf{Ignored stimulus} \\
    \midrule
    $1$ & Stream $1$-$1$ (Speaker A) & Stream $2$-$1$ (Speaker B) \\
    $2$ & Stream $1$-$2$ (Speaker A) & Stream $2$-$2$ (Speaker B) \\
    $3$ & Stream $3$-$1$ (Speaker C) & Stream $4$-$1$ (Speaker B) \\
    $4$ & Stream $3$-$2$ (Speaker C) & Stream $4$-$2$ (Speaker B) \\
    $5$ & Stream $2$-$1$ (Speaker B) & Stream $1$-$1$ (Speaker A) \\
    $6$ & Stream $2$-$2$ (Speaker B) & Stream $1$-$2$ (Speaker A) \\
    $7$ & Stream $4$-$1$ (Speaker B) & Stream $3$-$1$ (Speaker C) \\
    $8$ & Stream $4$-$2$ (Speaker B) & Stream $3$-$2$ (Speaker C) \\
    \bottomrule
  \end{tabular}
  }
\end{table}

\subsection{Within-trial vs. cross-trial performance}
Figure~\ref{fig:fig1b} presents within-trial and cross-trial performance under the protocols defined above. We report SI-SDR between the extracted signal and the ground-truth attended speech, and attended-speaker selection accuracy defined as the fraction of segments in which the extracted signal achieves higher SI-SDR with respect to the attended speech than with the ignored speech. While within-trial evaluation yields a median accuracy close to 90\%, cross-trial evaluation falls below the chance level, indicating that the model fails to reliably identify the attended speaker in unseen trials. Notably, in cross-trial evaluation we occasionally observe strongly negative SI-SDR values. Since SI-SDR is computed with respect to the attended reference, extracting the ignored speaker, as reflected in low selection accuracy, naturally yields very low SI-SDR even when the extracted output is perceptually clean. Cross-trial results also show substantial variance across the four folds and random seeds, emphasizing that statistically careful reporting is necessary.

\subsection{Linear probing of trial information}
NeuroHeed takes a speech mixture and a concurrent EEG segment as input, mapping the EEG through an encoder to a conditioning embedding that guides a separation network to produce an estimate of the attended speech. The entire system is optimized end-to-end to maximize the SI-SDR between the estimated waveform and the ground-truth attended speech.

However, since the attended speaker is fixed within each trial, the trial index deterministically specifies the target speaker. Thus, if trial identity is recoverable from EEG, it provides a shortcut under within-trial splits. We therefore quantify how easily trial identity is recoverable from the learned EEG embeddings via linear probing. Specifically, we first train NeuroHeed under the within-trial protocol and then freeze the learned EEG encoder. Using the frozen encoder to extract EEG embeddings, we train a linear classifier to predict the trial index among the 8 trials and evaluate this classifier on temporally disjoint held-out segments from the same trials. Because the probe is restricted to a linear model, strong trial classification indicates that trial-discriminative structure is readily accessible in the embedding space.

\begin{figure}[t]
  \centering
  \includegraphics[width=0.7\linewidth]{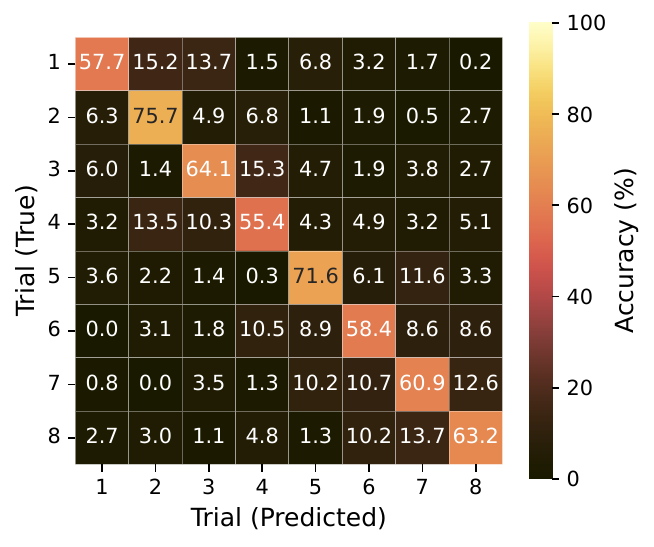}
  \caption{Confusion matrix of a linear probe predicting trial index from frozen EEG embeddings of an end-to-end trained NeuroHeed model on the KUL dataset.}
  \label{fig:probing}
\end{figure}

Figure~\ref{fig:probing} shows the results of the linear probe that predicts the trial index from frozen EEG embeddings. The predictions are strongly concentrated along the diagonal, with per-trial accuracies ranging from $55.4$\% to $75.7$\%, substantially above the $12.5$\% chance level for 8-way classification. This indicates that end-to-end training preserves significant trial-level structure in the EEG embeddings. Because such trial-linked information is non-transferable to unseen trials, it can inflate within-trial performance and help explain the pronounced degradation under cross-trial evaluation. This observation motivates the stress tests in the next subsection, which more directly examine whether extraction performance depends on trial identity.

\subsection{EEG--audio mismatch stress tests}
We next investigate whether the EEG-conditioned extraction performance under the within-trial protocol truly relies on temporally aligned EEG--audio correspondence, rather than trial-specific patterns. To this end, we conduct two EEG--audio mismatch stress tests.

\subsubsection{Test-time EEG shuffle}
Our first stress test perturbs EEG only at test time while keeping the trained model fixed. Specifically, we evaluate a NeuroHeed model trained under the standard within-trial protocol, but replace each test EEG segment with another EEG segment drawn from the same trial, at least 30 seconds away from the original time index. This manipulation breaks segment-level EEG--audio alignment while preserving trial identity. If within-trial performance were primarily driven by attention signatures temporally aligned to the concurrent speech, this mismatch should substantially degrade the performance. Instead, Table~\ref{tab:mismatch} shows that within-trial performance remains essentially unchanged under this perturbation. This invariance suggests that high within-trial scores are not driven primarily by temporally aligned, segment-specific EEG attention information.

\begin{figure*}[t]
  \centering
  \includegraphics[width=\textwidth]{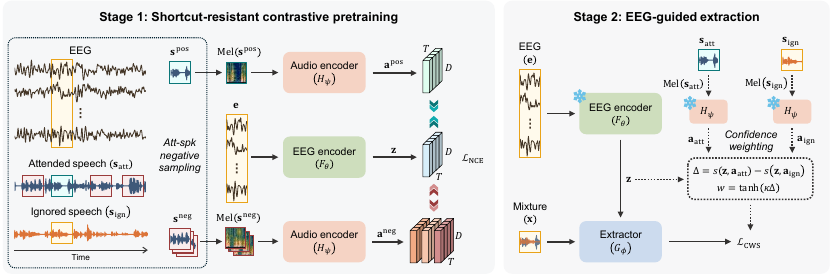}
  \caption{Overview of TRUST-TSE. Stage~1 trains an EEG encoder via cross-modal contrastive pretraining with attended-speaker negative sampling to suppress trial-identity shortcuts and produce attention-relevant EEG embeddings. Stage~2 trains an EEG-conditioned speech extractor with the EEG encoder frozen, using a confidence-weighted SI-SDR objective based on EEG--source similarity.}
  \label{fig:overview}
\end{figure*}

\subsubsection{Trial-wise permutation of EEG--audio pairing}
The second stress test disrupts EEG--audio correspondence already during training, while intentionally preserving the trial-level structure that can enable shortcut learning. We create a mismatched dataset by permuting EEG recordings across trials and pairing the EEG from trial $i$ with the audio mixture from a different trial $j$, which breaks both temporal alignment and the mapping between EEG and the attended speaker identity in the paired mixture. We then train NeuroHeed end-to-end on this perturbed dataset and evaluate it under the same within-trial protocol using the permuted pairing in the corresponding validation and test splits. Surprisingly, Table~\ref{tab:mismatch} shows no degradation in within-trial performance. This result demonstrates that the model can obtain strong within-trial scores even when the EEG provides no valid information about the concurrently presented speech. Notably, in the KUL dataset the same audio mixture appears in two trials with opposite attended labels (Table~\ref{tab:kul}), and our trial-wise permutation preserves this property, ensuring that the target is not identifiable from audio alone. Under such conditions, a plausible mechanism is that the model exploits trial-specific EEG structure as an identifier and learns a consistent trial-to-speaker selection rule that generalizes across temporally disjoint segments within the same trial, yet fails when evaluated on unseen trials.

\begin{table}[t]
  \caption{Within-trial performance of NeuroHeed under standard evaluation and two EEG--audio mismatch stress tests. The stress-test results are reported as mean $\pm$ standard deviation over 4 within-trial folds and two random seeds (8 runs in total).}
  \label{tab:mismatch}
  \centering
  \footnotesize
  \begin{tabular}{rccc}
    \toprule
     & \shortstack{\textbf{Standard}\\~} & \shortstack{\textbf{Test-time}\\\textbf{shuffle}} & \shortstack{\textbf{Trial-wise}\\\textbf{permutation}} \\
    \midrule
    SI-SDR (dB) & 12.12 $\pm$ 1.25 & 12.01 $\pm$ 1.33 & 12.26 $\pm$ 0.60 \\
    Accuracy (\%) & 86.90 $\pm$ 2.29 & 86.60 $\pm$ 2.29 & 86.81 $\pm$ 1.33 \\
    \bottomrule
  \end{tabular}
\end{table}

Together with the marked within- vs.\ cross-trial performance gap and the strong linear decodability of trial identity from end-to-end EEG embeddings, these mismatch tests show that high within-trial scores alone do not indicate temporally aligned EEG-based attention tracking. This underscores the limitations of within-trial evaluation as an incomplete reliability measure and motivates training procedures that discourage trial-linked shortcuts and improve robustness to unseen trials, as pursued by our proposed method in the next section.

\section{TRUST-TSE: trial-robust EEG-guided target speech extraction}
\subsection{Overview}

We propose TRUST-TSE, a two-stage training framework for trial-robust EEG-guided TSE that prevents shortcut learning driven by trial-specific EEG patterns. Given an EEG segment $\mathbf{e}$ and a mixture waveform $\mathbf{x}$ (with corresponding clean sources $\mathbf{s}_\text{att}$ and $\mathbf{s}_\text{ign}$ during training), our goal is to estimate the attended speech $\hat{\mathbf{s}}_{\text{att}}$ on unseen trials. A direct end-to-end formulation must simultaneously separate the mixture and learn to use EEG-derived attention information to guide extraction, which makes optimization brittle and prone to shortcuts based on trial-specific EEG patterns. Therefore, TRUST-TSE decomposes the task into two distinct, more tractable subproblems and optimizes the framework in two stages (Figure~\ref{fig:overview}). In Stage~1, we pretrain an EEG encoder $F_{\theta}$ using a contrastive objective with attended-speaker negative sampling to produce an embedding $\mathbf{z}=F_{\theta}(\mathbf{e})$ that emphasizes attention-relevant structure while discouraging trial identity cues. In Stage~2, we train an extractor $G_{\phi}$ to perform extraction guided by the pretrained EEG representation, producing $\hat{\mathbf{s}}=G_{\phi}(\mathbf{x},\mathbf{z})$. We keep the EEG encoder $F_{\theta}$ fixed to preserve a stable conditioning representation and avoid reshaping it toward trial-specific shortcut cues. Crucially, Stage~2 uses a confidence-weighted SI-SDR objective, with confidence estimated from the pretrained EEG--audio similarity, to emphasize segments with clear guidance and down-weight ambiguous ones.

\subsection{Shortcut-resistant contrastive pretraining}
Stage~1 aims to learn an EEG representation that is informative for auditory attention while being robust to trial-specific nuisance structure. As diagnosed in Section~\ref{sec:diagnosing}, end-to-end optimization can encourage the EEG encoder to retain trial-discriminative patterns that become spuriously predictive under within-trial splits. We therefore pretrain the EEG encoder with a contrastive objective whose negative sampling is deliberately constructed to make such trial cues unhelpful.

\subsubsection{Cross-modal sequence embeddings}
For each training segment, we are given EEG $\mathbf{e}\in\mathbb{R}^{T_e\times C}$ and the clean source waveform $\mathbf{s}$. We pass EEG through an EEG encoder $F_\theta$ to obtain a temporally resolved embedding sequence $\mathbf{z}=F_\theta(\mathbf{e})\in\mathbb{R}^{T\times D}$. In parallel, we transform $\mathbf{s}$ into a mel spectrogram and feed it to an audio encoder $H_{\psi}$, producing an embedding sequence $\mathbf{a}=H_\psi(\mathrm{Mel}(\mathbf{s}))\in\mathbb{R}^{T_a\times D}$ in a shared embedding space. Because EEG and audio embeddings may have different temporal resolutions, we linearly interpolate $\mathbf{a}$ along the temporal dimension to match the length $T$ of $\mathbf{z}$. We $\ell_2$-normalize both $\mathbf{z}$ and $\mathbf{a}$ along the feature dimension so that dot products correspond to cosine similarity.

\subsubsection{Attended-speaker negative sampling}
A key design choice in Stage~1 is how to construct negative samples for contrastive learning. In cross-modal contrastive learning, a common choice is to use other samples in the minibatch as negatives~\cite{radford2021learning}. In our setting, however, we additionally aim to reduce sensitivity to trial-specific nuisance structure in EEG, which motivates a negative construction that controls for trial identity. Concretely, for each EEG segment we treat the temporally aligned attended speech segment as the positive, and we draw negatives from other, non-aligned speech segments spoken by the attended speaker within the same trial. For example, in a trial where the listener attends to speaker A while ignoring speaker B, both the positive and the negatives are segments uttered by speaker A. Consequently, even if the EEG encoder exploits trial-specific patterns to infer which trial an EEG segment comes from, this information alone cannot distinguish the positive from the negative speech segments, since all candidates share the same trial identity. This design thus encourages the encoder to resolve the contrastive task by capturing segment-level EEG--speech alignment, rather than by leveraging trial identity cues. The effect of this negative sampling strategy is examined in a controlled ablation (Section~\ref{sec:neg_abl}).

\subsubsection{Objective}
We train the EEG encoder $F_{\theta}$ and the audio encoder $H_{\psi}$ with a cross-modal contrastive objective. For each EEG segment $i$ in a minibatch, we form the positive pair $(\mathbf{z}_i,\mathbf{a}^{\text{pos}}_i)$ using the temporally aligned clean attended speech segment, and construct a set of negatives $\{\mathbf{a}^{\text{neg}}_j\}_{j\in\mathcal{N}(i)}$ via attended-speaker negative sampling as described above. We define the similarity score by averaging cosine similarity over time,
\begin{equation}
\label{eq:similarity}
s(\mathbf{z},\mathbf{a}) = \frac{1}{T}\sum_{t=1}^{T} \mathbf{z}_t^\top \mathbf{a}_t
\end{equation}
where $\mathbf{z}_t,\mathbf{a}_t\in\mathbb{R}^{D}$ are $\ell_2$-normalized embeddings at time step $t$. We then optimize a cross-modal contrastive loss~\cite{radford2021learning},
\begin{equation}
\label{eq:nce}
\mathcal{L}_{\text{NCE}}
= -\frac{1}{B}\sum_{i=1}^{B}\log
\frac{e^{s(\mathbf{z}_i,\mathbf{a}^{\text{pos}}_i)/\tau}}
{e^{s(\mathbf{z}_i,\mathbf{a}^{\text{pos}}_i)/\tau}+\sum_{j}{e^{s(\mathbf{z}_i,\mathbf{a}^{\text{neg}}_{j})/\tau}}}
\end{equation}
where $B$ is the batch size and $\tau$ is a temperature hyperparameter. Intuitively, this objective encourages the EEG encoder to produce embeddings that are preferentially aligned with the correct attended-speech segment, while reducing similarity to mismatched negative speech segments.

\subsection{EEG-guided extraction with confidence weighting}
Stage~2 trains an EEG-conditioned speech extractor $G_{\phi}$ to follow the EEG guidance provided by the pretrained representation learned in Stage~1. Concretely, given a mixture waveform $\mathbf{x}=\mathbf{s}_{\text{att}}+\mathbf{s}_{\text{ign}}$ and the corresponding EEG segment $\mathbf{e}$, we compute an EEG embedding sequence $\mathbf{z}=F_{\theta}(\mathbf{e})$ and estimate the target speech as $\hat{\mathbf{s}}=G_{\phi}(\mathbf{x},\mathbf{z})$. We keep $F_{\theta}$ frozen throughout Stage~2. As diagnosed in Section~\ref{sec:diagnosing}, under end-to-end training with a waveform-level SI-SDR objective, the model must simultaneously solve mixture separation and infer the target from EEG, which makes the intended conditioning pathway difficult to learn. By freezing $F_{\theta}$, we prevent the extractor optimization from drifting the EEG representation toward trial-specific shortcuts, and instead encourage $G_{\phi}$ to learn an explicit conditioning pathway that generalizes across trials.

At the same time, conditioning on a fixed embedding introduces a practical challenge. The segment-level conditioning signal provided by the fixed EEG embedding is not equally reliable for all segments. Depending on representation mismatch, EEG noise, or attention fluctuation, the embedding $\mathbf{z}$ can (i) strongly align with the attended source, (ii) provide little discriminative information, or (iii) appear more compatible with the ignored source for some segments. If the extractor is trained uniformly with a loss that always pushes $\hat{\mathbf{s}}$ toward $\mathbf{s}_{\text{att}}$ for all segments, then segments where the guide is ambiguous or contradictory can introduce conflicting gradients. In practice, this can make optimization unstable and encourage the extractor to reduce its dependence on EEG conditioning altogether, undermining the goal of Stage~2.

To address this, we introduce a confidence-weighted objective that modulates Stage~2 learning based on how clearly the frozen EEG embedding supports the trial-level supervision for each segment. During training, we have access to the clean sources $\mathbf{s}_{\text{att}}$ and $\mathbf{s}_{\text{ign}}$. Using the frozen audio encoder $H_{\psi}$ from Stage~1, we obtain their sequence embeddings $\mathbf{a}_{\text{att}}=H_{\psi}(\mathrm{Mel}(\mathbf{s}_{\text{att}}))$ and $\mathbf{a}_{\text{ign}}=H_{\psi}(\mathrm{Mel}(\mathbf{s}_{\text{ign}}))$, temporally aligned to $\mathbf{z}$. Using the similarity score $s(\cdot,\cdot)$ defined in Eq.~\ref{eq:similarity}, we form a similarity margin,
\begin{equation}
\label{eq:margin}
\Delta = s(\mathbf{z},\mathbf{a}_{\text{att}}) - s(\mathbf{z},\mathbf{a}_{\text{ign}}).
\end{equation}
We convert this margin into a bounded confidence weight as follows:
\begin{equation}
\label{eq:weight_signed}
w = \tanh(\kappa \Delta),
\end{equation}
where $\kappa>0$ controls the sharpness of weighting. The magnitude $|w|$ reflects how confidently the embedding differentiates the two sources (near $0$ when ambiguous, approaching $1$ when confident), and $w>0$ indicates that the EEG embedding is more compatible with the attended source than the ignored source for that segment, while $w<0$ indicates the opposite.

The key design question is how to train the extractor on segments with negative $w$, where the frozen conditioning embedding is more compatible with the ignored source than the attended source. A conservative option is to ignore negative weights by clamping them to zero, using only positive-weight segments where the guide supports the attended target. Alternatively, we interpret $w<0$ as indicating that, for these segments, the conditioning embedding is closer to the ignored source than the attended source. To encourage the extractor to remain responsive to the conditioning, we retain negative weights by using their magnitude as confidence and optimize a confidence-weighted SI-SDR objective:
\begin{equation}
\label{eq:cws}
\mathcal{L}_{\text{CWS}} = -\mathbb{E}\big[|w| \cdot \mathrm{SI\mbox{-}SDR}(\mathbf{s}^{\star}, \hat{\mathbf{s}})\big],
\end{equation}
where,
\begin{equation}
\mathbf{s}^{\star} = \begin{cases} \mathbf{s}_{\text{att}}, & w \ge 0,\\ \mathbf{s}_{\text{ign}}, & w < 0. \end{cases} \end{equation}
As a result, segments with strong negative weights contribute meaningful gradients rather than being dropped, and the extractor is explicitly trained to remain sensitive to the conditioning embedding. We report an ablation in Section~\ref{sec:confidence_ablation} comparing it to the positive-only variant. At test time, the extractor uses only $(\mathbf{x},\mathbf{e})$ to produce $\hat{\mathbf{s}}$. The similarity computation and confidence weighting are used only during training.

\begin{table*}[t]
  \caption{Cross-trial EEG-guided TSE results on the KUL and DTU datasets using 5\,s windows, comparing representative end-to-end baselines with our two-stage framework. Values are mean $\pm$ standard deviation across 8 runs (4 folds $\times$ 2 random seeds).}
  \label{tab:mainresult}
  \centering
  \footnotesize

  \begin{tabular}{c l c c c c}
    \toprule
    \textbf{Datasets} & \textbf{Methods}
    & \multicolumn{1}{c}{\textbf{Accuracy (\%)}}
    & \multicolumn{1}{c}{\textbf{SI-SDR$_{\text{att}}$-All (dB)}}
    & \multicolumn{1}{c}{\textbf{SI-SDR$_{\text{att}}$-Correct (dB)}}
    & \multicolumn{1}{c}{\textbf{SI-SDR$_{\text{ign}}$-Wrong (dB)}} \\
    \midrule

    \multirow{3}{*}{KUL}
      & NeuroHeed        & 37.56 $\pm$ 35.99 & $-$12.09 $\pm$ 14.66 & 9.52 $\pm$ 2.34 & 11.27 $\pm$ 3.58 \\
      & M3ANet           & 48.42 $\pm$ 9.41 & $-$0.19 $\pm$ 1.01 & 3.62 $\pm$ 0.28 & 3.58 $\pm$ 0.70 \\
      & TRUST-TSE (ours) & \bfseries 62.27 $\pm$ 2.51 & \bfseries 0.26 $\pm$ 1.58 & \bfseries 15.23 $\pm$ 0.76 & \bfseries 14.89 $\pm$ 1.73 \\
    \midrule

    \multirow{3}{*}{DTU}
      & NeuroHeed        & 55.79 $\pm$ 7.07 & 0.08 $\pm$ 0.50 & 10.40 $\pm$ 7.25 & 10.34 $\pm$ 7.35 \\
      & M3ANet           & 50.23 $\pm$ 1.54 & $-$0.57 $\pm$ 0.44 & 4.83 $\pm$ 0.54 & 4.95 $\pm$ 0.39 \\
      & TRUST-TSE (ours) & \bfseries 70.40 $\pm$ 1.42 & \bfseries 4.85 $\pm$ 0.95 & \bfseries 19.21 $\pm$ 0.53 & \bfseries 18.14 $\pm$ 0.64 \\
    \bottomrule
  \end{tabular}
\end{table*}

\subsection{Model architectures}
Our framework is compatible with a broad class of encoder and extractor architectures. To isolate the efficacy of the proposed training mechanism from architectural complexity, we employ standard, lightweight components in both stages. In Stage~1, we adopt simple convolutional encoders for $F_\theta$ and $H_\psi$ to produce sequence embeddings. Stage~2 adopts the time-domain architecture of NeuroHeed~\cite{pan2024neuroheed}, employing a Dual-Path RNN (DPRNN) separator~\cite{luo2020dual}. Specifically, conditioning is implemented by linearly interpolating the frozen EEG embedding to align temporally with the mixture feature sequence, followed by channel-wise concatenation.

\section{Experiments}
\label{experiments}
\subsection{Experimental setup}
\label{expsetup}
\subsubsection{Datasets}
We evaluate on two public auditory attention EEG datasets, KUL~\cite{biesmans2016auditory} and DTU~\cite{fuglsang2017noise}. Across all experiments, we use strict cross-trial splits.

The KUL dataset contains 64-channel EEG from 16 subjects recorded during a selective attention task with two concurrent speech streams presented to the left and right ears. We use 8 trials per subject, each approximately 6 minutes long. We follow the cross-trial protocol described in Section~\ref{subsection:kul} and report results using 4 folds with a 6:1:1 train/validation/test split at the trial level. We use the publicly released preprocessed recordings, which are high-pass filtered at 0.5 Hz, downsampled to 128 Hz, and denoised using multichannel Wiener filtering.

The DTU dataset comprises 64-channel EEG from 18 subjects attending to one of two concurrent Danish speech streams. Each trial has a duration of 50\,s, with sources positioned at $\pm 60^\circ$. We use the official preprocessed EEG data, with steps including 0.1\,Hz high-pass filtering, ocular artifact correction, and 64\,Hz downsampling. For each subject, we split the 60 trials into 45/7/8 train/validation/test trials, and report results over 4 folds. Since the attended target speaker is unchanged across consecutive trials in DTU, we hold out validation and test as a contiguous block of 15 trials rather than interleaving held-out trials, to reduce potential leakage from slow nonstationarities across adjacent trials.

Following prior studies~\cite{pan2024neuroheed,ijcai2025p894}, we primarily consider a known-subject setting. For each dataset, a single model is trained on data from all subjects and evaluated on the same subject pool. This setting reflects practical scenarios in which EEG recordings from the target users are available for training and the resulting model is deployed for those same users. We additionally evaluate unseen-subject generalization in the supplementary material (Section B.2). For both datasets, we segment continuous recordings into fixed-length samples using sliding windows of 2\,s, 5\,s, and 10\,s, with a 1\,s hop between consecutive windows.

\subsubsection{Training details}
We trained TRUST-TSE in two stages with different optimization settings. In Stage~1, we jointly optimized the EEG encoder $F_\theta$ and audio encoder $H_\psi$ for up to 50 epochs using AdamW with a learning rate of $5\times10^{-4}$ and a batch size of 64. We fixed the temperature $\tau$ in $\mathcal{L}_\text{NCE}$ to 0.07. In Stage~2, we froze $F_\theta$ and $H_\psi$ and trained the extractor $G_\phi$ for up to 100 epochs using AdamW with a learning rate of $3\times10^{-4}$ and a batch size of 8. We set the confidence scaling parameter to $\kappa=5$. We applied early stopping with a patience of 10 epochs based on the validation accuracy for Stage~1 and the validation loss $\mathcal{L}_{\text{CWS}}$ for Stage~2, and selected the best checkpoint in each stage.

\subsubsection{Evaluation metrics}
Following prior work~\cite{pan2024neuroheed,ijcai2025p894}, we evaluate extracted speech quality using objective metrics computed against the ground-truth attended speech, including SI-SDR~\cite{le2019sdr}, PESQ~\cite{rix2001perceptual}, and STOI~\cite{taal2011algorithm}. However, attended-reference scores alone can be hard to interpret in EEG-guided extraction, because they conflate target selection with separation quality. For example, a model may produce a clean estimate of the attended speech on correctly selected segments but a similarly clean estimate of the ignored speech on other segments, so averaging SI-SDR against the attended reference can appear artificially low despite strong separation quality. Accordingly, we report a set of complementary metrics that disentangle selection from signal quality, enabling an unambiguous interpretation of both aspects.

Let $\hat{\mathbf{s}}$ be the extracted waveform, and let $\mathbf{s}_{\text{att}}$ and $\mathbf{s}_{\text{ign}}$ be the attended and ignored clean references. We define a segment as correctly selected if $\mathrm{SI\mbox{-}SDR}(\hat{\mathbf{s}},\mathbf{s}_{\text{att}}) >
\mathrm{SI\mbox{-}SDR}(\hat{\mathbf{s}},\mathbf{s}_{\text{ign}})$.
We report in the main tables:
\begin{itemize}
\item \textbf{Accuracy}: the percentage of segments that are correctly selected.
\item \textbf{SI-SDR$_{\text{att}}$-All}: mean $\mathrm{SI\mbox{-}SDR}(\hat{\mathbf{s}},\mathbf{s}_{\text{att}})$ over all segments (the attended-reference metric commonly reported in prior work, enabling direct comparability).
\item \textbf{SI-SDR$_{\text{att}}$-Correct}: mean $\mathrm{SI\mbox{-}SDR}(\hat{\mathbf{s}},\mathbf{s}_{\text{att}})$ over correctly selected segments.
\item \textbf{SI-SDR$_{\text{ign}}$-Wrong}: mean $\mathrm{SI\mbox{-}SDR}(\hat{\mathbf{s}},\mathbf{s}_{\text{ign}})$ over incorrectly selected segments.
\end{itemize}
Together, these metrics make it straightforward to attribute performance differences to selection versus separation. For brevity, SI-SDR is the primary quality metric in the main text, and PESQ and STOI are reported in the supplementary material using the same (All / Correct / Wrong) decomposition.

\subsection{Main results}
Table~\ref{tab:mainresult} summarizes the main cross-trial results using 5\,s windows under the strict trial-level splits described in Section~\ref{expsetup}. As baselines, we consider NeuroHeed~\cite{pan2024neuroheed} and M3ANet~\cite{ijcai2025p894}, two recent EEG-guided TSE methods with publicly available code and strong reported performance. For each method, we first verified that the released implementation reproduces the performance reported in the original paper under its stated evaluation setup, and then retrained it on our strict cross-trial splits with identical preprocessing and evaluation metrics.

Across both datasets, TRUST-TSE yields the most reliable cross-trial target selection and the highest extracted speech quality against the attended reference. On KUL, TRUST-TSE achieves a selection accuracy of 62.27\%, outperforming M3ANet (48.42\%) and NeuroHeed (37.56\%). Consistently, on DTU, TRUST-TSE reaches 70.40\%, exceeding NeuroHeed (55.79\%) and M3ANet (50.23\%).

Importantly, SI-SDR$_{\text{att}}$-All should be interpreted jointly with selection accuracy, since the average over all segments can be low due to target selection errors, even when separation quality is high on correctly selected segments. For example, for NeuroHeed on DTU, the SI-SDR$_{\text{att}}$-All is 0.08\,dB. However, the extracted signal achieves an SI-SDR$_{\text{att}}$-Correct of 10.40\,dB on correctly selected segments and an SI-SDR$_{\text{ign}}$-Wrong of 10.34\,dB on incorrectly selected segments, indicating that errors are primarily due to target selection rather than separation. TRUST-TSE improves SI-SDR$_{\text{att}}$-All primarily by increasing the fraction of correctly selected segments, while also achieving high separation quality when selection is correct, with SI-SDR$_{\text{att}}$-Correct of 15.23\,dB on KUL and 19.21\,dB on DTU.

For clarity, we report the results using 5\,s windows in the main table. We provide full results in the supplementary material, including 2\,s and 10\,s windows as well as PESQ and STOI (Section B.1), and unseen-subject generalization results (Section B.2). The same trend is observed across these evaluations, with TRUST-TSE maintaining robust performance and consistently outperforming the baselines.

\subsection{Validating shortcut resistance in TRUST-TSE}

\begin{table}[t]
  \caption{Stage~2 selection accuracy (\%) of TRUST-TSE under standard evaluation and EEG–audio mismatch stress tests using 5\,s windows. Mean $\pm$ standard deviation over 8 runs.}
  \label{tab:stress_ours}
  \centering
  \footnotesize
  {\setlength{\tabcolsep}{10pt}
  \begin{tabular}{rccc}
    \toprule
     & \shortstack{\textbf{Standard}\\~} & \shortstack{\textbf{Test-time}\\\textbf{shuffle}} & \shortstack{\textbf{Trial-wise}\\\textbf{permutation}} \\
    \midrule
    KUL & 62.27 $\pm$ 2.51 & 44.96 $\pm$ 2.63 & 48.74 $\pm$ 1.06 \\
    DTU & 70.40 $\pm$ 1.42 & 49.51 $\pm$ 0.99 & 50.08 $\pm$ 1.14 \\
    \bottomrule
  \end{tabular}}
\end{table}

To verify that TRUST-TSE relies on meaningful EEG--audio correspondence rather than trial-linked shortcuts, we repeat the two mismatch stress tests used in Section~\ref{sec:diagnosing}. Table~\ref{tab:stress_ours} shows that, unlike end-to-end baselines, TRUST-TSE is not invariant to EEG--audio mismatch: both test-time EEG shuffle and trial-wise permutation cause substantial drops in Stage~2 selection accuracy on KUL (62.27\% to 44.96/48.74\%) and DTU (70.40\% to 49.51/50.08\%). The accuracies fall near the chance level, indicating that when segment-level EEG--speech alignment is disrupted, TRUST-TSE can no longer reliably select the attended speaker. Consistent with this, linear probing of the EEG embeddings shows that trial identity is no longer readily decodable on KUL, achieving near-chance accuracy (see supplementary material, Figure~S1) in contrast to the end-to-end NeuroHeed encoder. This behavior is in line with our objective of suppressing trial-linked shortcuts and encouraging extraction decisions to rely on aligned EEG evidence that transfers across trials.

\subsection{Ablations}

\subsubsection{Negative sampling for contrastive pretraining}
\label{sec:neg_abl}

\begin{table}[t]
  \caption{Cross-trial Stage~1 selection accuracy (\%) under different negative sampling strategies across window lengths. Values are mean $\pm$ standard deviation over 16 runs.}
  \label{tab:negative}
  \centering
  \footnotesize
  \begin{tabular}{crccc}
  \toprule
    & &
    \shortstack{\smash[b]{\textbf{Ignored}}\\\textbf{speaker}} &
    \shortstack{\textbf{In-batch}\\\textbf{(All speakers)}} &
    \shortstack{\textbf{Attended}\\\textbf{speaker}} \\
  \midrule
  \multirow{3}{*}{KUL}
    & 2\,s & 55.92 $\pm$ 4.80 & \bfseries 64.19 $\pm$ 6.03  &  59.57 $\pm$ 1.48 \\
    & 5\,s & 39.63 $\pm$ 14.99 & 64.05 $\pm$ 7.76 & \bfseries 65.62 $\pm$ 2.13 \\
    & 10\,s & 39.62 $\pm$ 17.65 & 65.76 $\pm$ 6.49 & \bfseries 67.25 $\pm$ 3.39 \\
  \midrule
  \multirow{3}{*}{DTU}
    & 2\,s & 59.05 $\pm$ 2.18 & 62.64 $\pm$ 1.32 & \bfseries 63.60 $\pm$ 1.02 \\
    & 5\,s & 64.60 $\pm$ 1.39 & 70.81 $\pm$ 1.12 & \bfseries 71.58 $\pm$ 0.88 \\
    & 10\,s & 67.39 $\pm$ 6.68 & 75.27 $\pm$ 0.44 & \bfseries 75.66 $\pm$ 0.61 \\
  \bottomrule
  \end{tabular}
\end{table}

To isolate the effect of the negative sampling strategy in Stage~1 contrastive pretraining, we compare three approaches while keeping all other components fixed. \textit{Ignored-speaker} samples negatives from speech segments uttered by the ignored speaker. \textit{In-batch} uses all other audio segments in the minibatch as negatives regardless of speaker identity, following the standard in-batch contrastive setup. \textit{Attended-speaker} (ours) samples negatives from non-aligned speech segments uttered by the attended speaker within the same trial.

Since this ablation focuses on contrastive pretraining, we evaluate the pretrained encoders using a Stage~1 selection accuracy defined directly from the cross-modal similarity scores $s(\cdot,\cdot)$ (Eq.~\ref{eq:similarity}). For each segment, we compute $\mathbf{z}=F_{\theta}(\mathbf{e})$ and $\mathbf{a}_{\text{att}},\mathbf{a}_{\text{ign}}$ from the clean sources using $H_{\psi}$. A segment is counted as correctly selected if $s(\mathbf{z},\mathbf{a}_{\text{att}}) > s(\mathbf{z},\mathbf{a}_{\text{ign}})$, and we report the resulting accuracy under the cross-trial splits described in Section~\ref{expsetup}, averaged over 4 folds and 4 random seeds.

Table~\ref{tab:negative} shows that \textit{Attended-speaker} negatives consistently yield the most robust performance, combining high selection accuracy with notably low variance. By contrast, \textit{Ignored-speaker} negatives result in the weakest and least reliable embeddings, failing to transfer robustly across unseen trials. While \textit{In-batch} negatives achieve competitive mean accuracy, they exhibit substantially higher variance, particularly on KUL (e.g., at 5\,s, 64.05 $\pm$ 7.76\% with a wide 54.09--81.67\% range over 16 runs). This instability is further reflected in the less consistent gains with increasing window length on KUL, in contrast to the monotonic gains of \textit{Attended-speaker}.

These results support the motivation for attended-speaker negative sampling. When negatives come exclusively from the ignored source, the contrastive task can be solved using coarse trial-specific patterns that separate attended from ignored streams, without requiring fine-grained, temporally localized EEG--speech alignment, which corresponds to the failure mode diagnosed in Section~3. In contrast, attended-speaker negatives remove this easy separation by constructing positives and negatives that share the attended speaker and trial, thereby encouraging the EEG encoder to rely on segment-level alignment signals that transfer across trials. In-batch sampling mixes heterogeneous negatives (including both easy and difficult cases), which can lead to less consistent optimization and larger variability.

\subsubsection{Confidence weighting}
\label{sec:confidence_ablation}

\begin{table}[t]
  \captionsetup{skip=6pt}
  \caption{Cross-trial Stage~2 selection accuracy for confidence-weighting variants on the KUL and DTU datasets. Results are reported as mean $\pm$ standard deviation across 8 runs.}
  \label{tab:confidence}
  \centering
  \footnotesize
  {\setlength{\tabcolsep}{4pt}
  \begin{tabular}{ccccc}
  \toprule
    & 
    \shortstack{\textbf{No CW}\\~} &
    \shortstack{\smash[b]{\textbf{Pos-only}}\\\textbf{tanh CW}} &
    \shortstack{\smash[b]{\textbf{Pos+neg}}\\\textbf{tanh CW}} &
    \shortstack{\smash[b]{\textbf{Pos+neg}}\\\textbf{sig CW}} \\
  \midrule
  KUL
    & 39.07 $\pm$ 9.97 & 42.73 $\pm$ 9.05 &\bfseries 62.27 $\pm$ 2.51 & 45.30 $\pm$ 3.69 \\
  DTU
    & 65.46 $\pm$ 2.66 & 70.15 $\pm$ 1.00 & \bfseries 70.40 $\pm$ 1.42 & 67.38 $\pm$ 2.89 \\
  \bottomrule
  \end{tabular}}
\end{table}

We examine the confidence-weighted SI-SDR objective used in Stage~2, focusing on handling segments with negative similarity margins ($\Delta<0$). We compare four weighting policies using identical pretrained encoders. \textit{No CW} removes confidence weighting and always uses the attended reference $\mathbf{s}_{\text{att}}$. \textit{Pos-only tanh CW} applies $w=\tanh(\kappa\Delta)$ but clamps negative weights to zero, learning only from segments where the embedding favors the attended source. \textit{Pos+neg tanh CW} (ours) retains negative weights by using their magnitude $|w|$ as confidence and selecting a reference $\mathbf{s}^{\star}$ consistent with the sign of $w$ (Eq.~\ref{eq:cws}). Finally, \textit{Pos+neg sig CW} retains $\Delta<0$ segments without reference switching: it uses a sigmoid weight $w=\sigma(\kappa\Delta)\in(0,1)$ to down-weight segments with conflicting guidance while still training with $\mathbf{s}_{\text{att}}$ as the reference.

Table~\ref{tab:confidence} reports cross-trial Stage~2 selection accuracy under the same setup as Table~\ref{tab:mainresult}. Confidence weighting is beneficial overall, as \textit{No CW} underperforms the weighted variants. On KUL, \textit{Pos+neg tanh CW} clearly outperforms both \textit{Pos-only tanh CW} and \textit{Pos+neg sig CW} (62.27\% vs.\ 42.73/45.30\%), whereas on DTU, the performance gap among the weighted variants is much smaller. This disparity aligns with the higher incidence of negative margins on KUL ($P(w<0)=0.256$) than on DTU ($0.168$) (see supplementary material, Section B.4). Discarding these segments (Pos-only) forfeits informative training data, while retaining them without resolving the conflict between the EEG embedding guidance and the training target (Sigmoid) risks confusing the extractor. Furthermore, since strong contradictions are rare in both datasets ($P(w<-0.5)=0.026$ on KUL and $0.022$ on DTU), switching references for high-confidence negative segments poses minimal risk of training destabilization.

The higher rate of negative margins on KUL likely reflects its greater difficulty in sustaining attention over prolonged trials: same-gender speakers and long trials ($\sim$6\,min) compared to mixed-gender and short trials (50\,s) in DTU. These factors may increase segment-level disagreements between the frozen embedding and the trial label. We note, however, that without instantaneous attention ground truth, we cannot definitively attribute negative weights to genuine attention switches.

\subsubsection{Contrastive EEG embeddings vs. envelope decoding}

Stage~1 in TRUST-TSE learns an EEG representation that supports target selection by comparing its cross-modal similarity to the candidate sources. A natural alternative is to substitute our contrastive EEG latent with EEG-to-speech amplitude envelope reconstruction, a long-standing standard in AAD that emphasizes slow temporal dynamics and contains minimal speaker-specific information. Since envelope decoding is expected to be insensitive to trial-identity shortcuts, we evaluate it as a competing Stage~1 design under the same cross-trial protocol.

We compare mTRF~\cite{crosse2016mtrf} and two recent deep neural envelope decoders (VLAAI~\cite{accou2023decoding} and ADTNet~\cite{liu2024adt}). For all methods, hyperparameters are tuned to maximize validation Stage~1 selection accuracy (see supplementary material, Section B.6). In addition, to decouple architectural capacity from the learning target, we include an architecture-matched variant that uses the same EEG encoder as TRUST-TSE but replaces the contrastive objective with an envelope reconstruction objective using the negative Pearson correlation loss, matching VLAAI and ADTNet (TRUST-TSE (envelope)). Following Section~\ref{sec:neg_abl}, we evaluate all methods by Stage~1 selection accuracy. The scoring metric depends on the representation domain: latent cosine similarity (Eq.~\ref{eq:similarity}) for TRUST-TSE (latent) and Pearson correlation of reconstructed envelopes for the baselines.

Table~\ref{tab:envelope} shows that the contrastive latent is consistently more discriminative than envelope-based signals in Stage~1 selection accuracy. On KUL, TRUST-TSE (latent) achieves 65.62\%, whereas mTRF and TRUST-TSE (envelope) reach 60.96\% and 60.28\%, respectively. On DTU, TRUST-TSE (latent) attains 71.58\%, exceeding 64.49\% with mTRF and 63.96\% with TRUST-TSE (envelope). The architecture-matched envelope reconstruction variant follows the envelope baselines rather than the latent, indicating that the gap is driven by the Stage~1 training objective rather than the encoder architecture. Overall, envelope decoding provides a non-negligible signal for target selection, but the contrastive latent achieves superior selection accuracy.

\section{Conclusion}
This work reexamines the state of EEG-guided TSE, revealing that exceptional within-trial performance can stem from trial-linked shortcuts rather than genuine neural decoding. We address this challenge with TRUST-TSE, a two-stage framework that learns trial-robust representations via contrastive alignment and guides extraction using confidence-aware supervision. Across strict cross-trial evaluations, TRUST-TSE improves reliability on unseen trials compared to end-to-end baselines. Ultimately, these findings underscore the necessity of moving beyond permissive protocols toward methods and evaluations that support robust, practical neuro-steered hearing technologies.

\begin{table}[t]
  \captionsetup{skip=6pt}
  \caption{Cross-trial Stage~1 selection accuracy (\%) for envelope baselines and TRUST-TSE variants with 5\,s windows (mean $\pm$ standard deviation over 16 runs).}
  \label{tab:envelope}
  \centering
  \footnotesize
  {\setlength{\tabcolsep}{10pt}
  \renewcommand{\arraystretch}{0.95}
  \begin{tabular}{lcc}
  \toprule
    & \textbf{KUL} & \textbf{DTU} \\
  \midrule
  mTRF
    & 60.96 $\pm$ 0.81 & 64.49 $\pm$ 0.77 \\
  VLAAI
    & 58.33 $\pm$ 1.08 & 63.13 $\pm$ 0.75 \\
  ADTNet
    & 54.83 $\pm$ 1.91 & 62.59 $\pm$ 0.99 \\
  TRUST-TSE (envelope)
    & 60.28 $\pm$ 1.33 & 63.96 $\pm$ 1.08 \\
  TRUST-TSE (latent)
    & \bfseries 65.62 $\pm$ 2.13 & \bfseries 71.58 $\pm$ 0.88 \\
  \bottomrule
  \end{tabular}}
\end{table}

\section{Limitations}
While TRUST-TSE improves cross-trial generalization, our evaluation is limited to publicly available EEG datasets that are modest in scale and diversity. Larger and more heterogeneous multi-session recordings could enable more stable attention-related representations for both end-to-end and two-stage approaches. However, trial-linked shortcuts and evaluation leakage may persist even with higher model capacity, making strict cross-trial evaluation and shortcut-aware training essential for assessing progress. Consequently, our results highlight both a methodological direction and a critical need for deployment-oriented datasets that reduce reliance on trial-specific cues.

\section{Acknowledgments}
This work was partly supported by the National Research Foundation of Korea (NRF) grant funded by the Korea government (MSIT) [No. RS-2024-00461617, 50\%], Institute of Information \& communications Technology Planning \& Evaluation (IITP) grant funded by the Korea government (MSIT) [No. RS-2022-II220320, 2022-0-00320, 45\%] and [No. RS-2021-II211343, 5\%].

\section{Generative AI use disclosure}
We utilized ChatGPT 5.2 for the purpose of grammatical error correction and polishing the manuscript. The authors are accountable for the work and content of the paper.

\bibliographystyle{IEEEtran}
\bibliography{mybib}

\clearpage
\input{supp}

\end{document}

%% file: supp.tex
\appendix
\begin{center}
\textbf{\LARGE Supplementary Material}\\
\end{center}

\renewcommand{\thefigure}{S\arabic{figure}}
\renewcommand{\thetable}{S\arabic{table}}

\setcounter{figure}{0}
\setcounter{table}{0}

\section{Implementation details}
\subsection{Model architectures}
Tables~\ref{tab:eeg_encoder} and~\ref{tab:mel_encoder} summarize the block-level architectures of the EEG encoder $F_{\theta}$ and the audio encoder $H_{\psi}$ to produce cross-modal sequence embeddings. For Stage~1 and the confidence computation in Stage~2, we convert each clean source waveform $\mathbf{s}$ (8~kHz, mono) to a log-mel spectrogram $\mathrm{Mel}(\mathbf{s})$ with 80 mel bins using a short-time Fourier transform (STFT) window of 25~ms and hop of 10~ms (FFT size $n_{\mathrm{fft}}{=}200$, power=2.0). Finally, we convert it to a dB scale before feeding it to $H_{\psi}$.

In Stage~2, we adopt the time-domain NeuroHeed extractor with a DPRNN separator. Given a mixture waveform $\mathbf{x}$, a single 1-D convolutional layer (kernel 20, stride 10) produces a sequence of mixture features. The EEG embedding $\mathbf{z}$ is linearly interpolated along the time axis to match the temporal length of the mixture features and concatenated along the feature dimension before DPRNN segmentation, followed by a $1{\times}1$ projection. In addition, we apply the same concatenation after each DPRNN block using the segmented EEG features, again followed by a $1{\times}1$ projection. For the DPRNN separator, we follow the default NeuroHeed configuration with hidden size $H=128$, chunk length $K=100$, and $R=6$ dual-path blocks. The DPRNN separator estimates a time-domain mask, which is applied to the mixture features and reconstructed by a learned linear synthesis transform followed by overlap-and-add.

\begin{table}[h]
  \caption{EEG encoder $F_{\theta}$ architecture.
  Conv1D operates on the temporal axis.}
  \label{tab:eeg_encoder}
  \centering
  \footnotesize
  {\setlength{\tabcolsep}{6pt}
  \begin{tabular}{c l l}
    \toprule
    \textbf{Block} & \textbf{Operation} & \textbf{Block parameters} \\
    \midrule
    \makecell{Channel\\projection} & PWConv1D & $C{=}48$, $k{=}1$ \\
    \midrule
    \multirow{6}{*}{\makecell{Depthwise\\convolution\\block\\(x4)}}
      & Right-only pad & $padR=d(k-1)$  \\
      & DWConv1D & $C{=}48$, $k{=}15$, $d\in\{1, 2, 4, 8\}$  \\
      & GroupNorm & $\text{groups}=1$ \\
      & Activation & SiLU \\
      & PWConv1D & $C{=}48$, $k{=}1$ \\
      & Dropout + skip & $p{=}0.15$ \\
    \bottomrule
  \end{tabular}}
\end{table}

\begin{table}[h]
  \caption{Audio encoder $H_{\psi}$ architecture.
  Conv2D operates on the time--frequency plane.}
  \label{tab:mel_encoder}
  \centering
  \footnotesize
  {\setlength{\tabcolsep}{3pt}
  \begin{tabular}{c l l}
    \toprule
    \textbf{Block} & \textbf{Operation} & \textbf{Block parameters} \\
    \midrule

    \multirow{4}{*}{\makecell{Spectral\\convolution\\block\\(x3)}}
      & Conv2D & $C\in\{32, 64, 128\}$, $k{=}3$, $pad{=}1$ \\
      & GroupNorm & $\text{groups}=C/4$ \\
      & Activation & GELU \\
      & MaxPool2D & $(2,1)$ (halve $F$) \\
    \midrule

     \makecell{Frequency\\pooling} & MeanPool & over $F$ \\
    \midrule

    \multirow{6}{*}{\makecell{Temporal\\convolution\\block\\(x2)}}
      & Conv1D & $C{=}128$, $k{=}5$, $d\in\{1,2\}$, $pad{=}2d$ \\
      & BatchNorm & -- \\
      & Activation & GELU \\
      & PWConv1D & $C{=}128$, $k{=}1$ \\
      & BatchNorm & -- \\
      & Skip (add) + act & GELU \\
    \midrule

   \multirow{3}{*}{\makecell{Linear\\projection}} & Linear & $D{=}48$ \\
   & Activation & GELU \\
   & LayerNorm & -- \\
    \bottomrule
  \end{tabular}}
\end{table}

\section{Additional experimental results}
\subsection{Main results: full table}
Table~\ref{tab:full_main} reports the full cross-trial results of on KUL and DTU for three window lengths (2/5/10\,s), including PESQ and STOI in addition to SI-SDR. We follow the cross-trial folds described in Section~5.1 (known-subject setting). Following Section~5.1.3, we decompose each metric into All$_{\text{att}}$, Correct$_{\text{att}}$, and Wrong$_{\text{ign}}$ to disentangle target selection from separation quality. All$_{\text{att}}$ should be interpreted jointly with accuracy, since the attended-reference average can be low when the system cleanly extracts the ignored speaker on mis-selected segments.

\begin{table*}[t]
\caption{Full main results with cross-trial evaluation for three EEG window lengths (2/5/10\,s). Values are mean$\pm$std over 8 runs (4 folds $\times$ 2 random seeds).}
\centering
\scriptsize
\setlength{\tabcolsep}{4pt}

\begin{tabular}{rl c ccc ccc ccc}
\toprule
\multicolumn{2}{c}{} & \multicolumn{1}{c}{\textbf{Accuracy}} &
\multicolumn{3}{c}{\textbf{SI-SDR (dB)}} & \multicolumn{3}{c}{\textbf{PESQ}} & \multicolumn{3}{c}{\textbf{STOI}} \\
\cmidrule(lr){4-6}\cmidrule(lr){7-9}\cmidrule(lr){10-12}
 & & \textbf{(\%)} & \textbf{All$_{\text{att}}$} & \textbf{Correct$_{\text{att}}$} & \textbf{Wrong$_{\text{ign}}$} & \textbf{All$_{\text{att}}$} & \textbf{Correct$_{\text{att}}$} & \textbf{Wrong$_{\text{ign}}$} & \textbf{All$_{\text{att}}$} & \textbf{Correct$_{\text{att}}$} & \textbf{Wrong$_{\text{ign}}$} \\
\midrule

\multicolumn{12}{c}{\textbf{KUL dataset}}\\
\midrule
\multirow{3}{*}{2s}  & NeuroHeed & 37.85$\pm$34.22 & -10.52$\pm$13.05 & 9.11$\pm$2.40 & 10.89$\pm$3.41 & 1.56$\pm$0.32 & 2.19$\pm$0.20 & 2.27$\pm$0.15 & 0.48$\pm$0.22 & 0.82$\pm$0.05 & 0.85$\pm$0.03 \\
                     & M3ANet    & 48.19$\pm$6.97 & \bfseries -0.07$\pm$0.99 & 4.53$\pm$0.25 & 4.17$\pm$0.78 & 1.71$\pm$0.06 & 1.97$\pm$0.13 & 2.20$\pm$0.10 & \bfseries 0.71$\pm$0.02 & 0.81$\pm$0.03 & 0.82$\pm$0.02 \\
                     & TRUST-TSE & \bfseries 55.41$\pm$3.47 & -2.02$\pm$0.90 & \bfseries 13.62$\pm$1.20 & \bfseries 13.19$\pm$1.73 & \bfseries 1.94$\pm$0.16 & \bfseries 2.50$\pm$0.22 & \bfseries 2.73$\pm$0.21 & 0.63$\pm$0.03 & \bfseries 0.88$\pm$0.03 & \bfseries 0.89$\pm$0.02 \\
\addlinespace[4pt]
\multirow{3}{*}{5s}  & NeuroHeed & 37.56$\pm$35.99 & -12.09$\pm$14.66 & 9.52$\pm$2.34 & 11.27$\pm$3.58 & 1.57$\pm$0.33 & 2.24$\pm$0.24 & 2.29$\pm$0.18 & 0.46$\pm$0.25 & 0.84$\pm$0.05 & 0.86$\pm$0.04 \\
                     & M3ANet    & 48.42$\pm$9.41 & -0.19$\pm$1.01 & 3.62$\pm$0.28 & 3.58$\pm$0.70 & 1.69$\pm$0.07 & 1.90$\pm$0.12 & 2.14$\pm$0.11 & \bfseries 0.71$\pm$0.02 & 0.80$\pm$0.03 & 0.81$\pm$0.02 \\
                     & TRUST-TSE & \bfseries 62.27$\pm$2.51 & \bfseries 0.26$\pm$1.58 & \bfseries 15.23$\pm$0.76 & \bfseries 14.89$\pm$1.73 & \bfseries 2.14$\pm$0.14 & \bfseries 2.67$\pm$0.16 & \bfseries 2.91$\pm$0.20 & 0.67$\pm$0.03 & \bfseries 0.91$\pm$0.02 & \bfseries 0.91$\pm$0.01 \\
\addlinespace[4pt]
\multirow{3}{*}{10s} & NeuroHeed & 36.09$\pm$38.19 & -12.49$\pm$15.09 & 9.29$\pm$2.62 & 11.47$\pm$3.26 & 1.58$\pm$0.33 & 2.23$\pm$0.28 & 2.31$\pm$0.20 & 0.45$\pm$0.26 & 0.84$\pm$0.05 & 0.86$\pm$0.03 \\
                     & M3ANet    & 46.47$\pm$10.34 & -0.33$\pm$1.11 & 3.04$\pm$0.32 & 3.13$\pm$0.78 & 1.68$\pm$0.08 & 1.86$\pm$0.11 & 2.11$\pm$0.11 & \bfseries 0.72$\pm$0.02 & 0.79$\pm$0.03 & 0.80$\pm$0.02 \\
                     & TRUST-TSE & \bfseries 65.05$\pm$2.70 & \bfseries 1.29$\pm$0.89 & \bfseries 14.76$\pm$0.68 & \bfseries 14.42$\pm$1.23 & \bfseries 2.14$\pm$0.17 & \bfseries 2.59$\pm$0.21 & \bfseries 2.84$\pm$0.19 & 0.68$\pm$0.03 & \bfseries 0.90$\pm$0.02 & \bfseries 0.91$\pm$0.01 \\
\midrule

\multicolumn{12}{c}{\textbf{DTU dataset}}\\
\midrule
\multirow{3}{*}{2s}  & NeuroHeed & 49.99$\pm$1.67 & -1.03$\pm$0.64 & 7.04$\pm$0.41 & 7.10$\pm$0.18 & 1.80$\pm$0.07 & 2.10$\pm$0.12 & 2.10$\pm$0.09 & 0.67$\pm$0.02 & 0.81$\pm$0.01 & 0.81$\pm$0.01 \\
                     & M3ANet    & 49.87$\pm$1.68 & -1.25$\pm$0.64 & 6.94$\pm$0.48 & 6.98$\pm$0.29 & 1.91$\pm$0.06 & 2.30$\pm$0.09 & 2.29$\pm$0.06 & \bfseries 0.68$\pm$0.01 & 0.83$\pm$0.01 & 0.82$\pm$0.01 \\
                     & TRUST-TSE & \bfseries 61.17$\pm$1.12 & \bfseries -0.23$\pm$0.67 & \bfseries 15.20$\pm$0.51 & \bfseries 14.10$\pm$0.68 & \bfseries 2.24$\pm$0.03 & \bfseries 2.81$\pm$0.03 & \bfseries 2.79$\pm$0.06 & 0.66$\pm$0.01 & \bfseries 0.86$\pm$0.01 & \bfseries 0.85$\pm$0.01 \\
\addlinespace[4pt]
\multirow{3}{*}{5s}  & NeuroHeed & 55.79$\pm$7.07 & 0.08$\pm$0.50 & 10.40$\pm$7.25 & 10.34$\pm$7.35 & 2.08$\pm$0.31 & 2.43$\pm$0.52 & 2.44$\pm$0.54 & 0.69$\pm$0.03 & 0.84$\pm$0.04 & 0.83$\pm$0.05 \\
                     & M3ANet    & 50.23$\pm$1.54 & -0.57$\pm$0.44 & 4.83$\pm$0.54 & 4.95$\pm$0.39 & 1.90$\pm$0.05 & 2.14$\pm$0.09 & 2.15$\pm$0.09 & 0.72$\pm$0.01 & 0.81$\pm$0.01 & 0.81$\pm$0.01 \\
                     & TRUST-TSE & \bfseries 70.40$\pm$1.42 & \bfseries 4.85$\pm$0.95 & \bfseries 19.21$\pm$0.53 & \bfseries 18.14$\pm$0.64 & \bfseries 2.63$\pm$0.05 & \bfseries 3.15$\pm$0.06 & \bfseries 3.10$\pm$0.08 & \bfseries 0.73$\pm$0.01 & \bfseries 0.90$\pm$0.01 & \bfseries 0.89$\pm$0.01 \\
\addlinespace[4pt]
\multirow{3}{*}{10s} & NeuroHeed & 51.25$\pm$4.31 & -0.41$\pm$0.49 & 5.78$\pm$5.75 & 5.88$\pm$5.86 & 1.89$\pm$0.22 & 2.07$\pm$0.39 & 2.09$\pm$0.40 & 0.70$\pm$0.03 & 0.79$\pm$0.05 & 0.79$\pm$0.05 \\
                     & M3ANet    & 50.36$\pm$1.73 & -0.31$\pm$0.37 & 3.56$\pm$0.55 & 3.64$\pm$0.34 & 1.94$\pm$0.06 & 2.11$\pm$0.10 & 2.11$\pm$0.06 & 0.74$\pm$0.01 & 0.81$\pm$0.02 & 0.81$\pm$0.01 \\
                     & TRUST-TSE & \bfseries 74.76$\pm$1.50 & \bfseries 7.05$\pm$0.94 & \bfseries 21.17$\pm$0.69 & \bfseries 21.57$\pm$3.19 & \bfseries 2.87$\pm$0.07 & \bfseries 3.34$\pm$0.07 & \bfseries 3.27$\pm$0.08 & \bfseries 0.76$\pm$0.01 & \bfseries 0.93$\pm$0.02 & \bfseries 0.92$\pm$0.01 \\

\bottomrule
\end{tabular}
\label{tab:full_main}
\end{table*}

\subsection{Generalization to unseen subjects}

Following prior studies, our main experiments adopt a known-subject setting, where a single model is trained on data from all subjects and evaluated on the same subject pool. In some deployment scenarios, however, EEG recordings from a target user may be unavailable during training. We therefore additionally evaluate generalization to unseen subjects by training the model on a subset of subjects and testing it on held-out subjects that are never observed during training. Since all trials from the test subjects are unseen during training, this evaluation also corresponds to the cross-trial regime. We construct four cross-subject folds with disjoint validation and test subject sets.  For KUL, each fold selects four held-out subjects from the full set of 16, with two used for validation and two for testing. For DTU, we first remove two subjects arbitrarily to match the subject split ratio used for KUL and then hold out two subjects for validation and two for testing from the remaining 16 subjects. Concretely, we exclude subjects 17 and 18 according to the dataset subject indices without any subject-specific inspection. We use a 5\,s EEG window with a 1\,s hop size, following the same windowing procedure as in the main experiments.

Table~\ref{tab:unseen_sub} summarizes the results. Overall, TRUST-TSE maintains a clear advantage over end-to-end baselines in selection accuracy and separation quality.

\begin{table*}[t]
\caption{Target speech extraction performance on unseen subjects for KUL and DTU, reported as mean$\pm$std over 8 runs (4 folds $\times$ 2 random seeds).}
\centering
\scriptsize
\setlength{\tabcolsep}{4.3pt}

\begin{tabular}{rl c ccc ccc ccc}
\toprule
\multicolumn{2}{c}{} & \multicolumn{1}{c}{\textbf{Accuracy}} &
\multicolumn{3}{c}{\textbf{SI-SDR (dB)}} & \multicolumn{3}{c}{\textbf{PESQ}} & \multicolumn{3}{c}{\textbf{STOI}} \\
\cmidrule(lr){4-6}\cmidrule(lr){7-9}\cmidrule(lr){10-12}
 & & \textbf{(\%)} & \textbf{All$_{\text{att}}$} & \textbf{Correct$_{\text{att}}$} & \textbf{Wrong$_{\text{ign}}$} & \textbf{All$_{\text{att}}$} & \textbf{Correct$_{\text{att}}$} & \textbf{Wrong$_{\text{ign}}$} & \textbf{All$_{\text{att}}$} & \textbf{Correct$_{\text{att}}$} & \textbf{Wrong$_{\text{ign}}$} \\
\midrule

\multirow{3}{*}{KUL} & NeuroHeed & 50.04$\pm$0.05 & -0.04$\pm$0.11 & 3.80$\pm$0.42 & 3.81$\pm$0.43 & 1.78$\pm$0.01 & 2.01$\pm$0.02 & 2.01$\pm$0.02 & \bfseries 0.72$\pm$0.01 & 0.80$\pm$0.01 & 0.80$\pm$0.01 \\
                     & M3ANet    & 53.62$\pm$4.06 & 0.30$\pm$0.52 & 3.48$\pm$0.24 & 3.19$\pm$0.29 & 1.70$\pm$0.07 & 1.88$\pm$0.10 & 2.01$\pm$0.04 & 0.71$\pm$0.01 & 0.79$\pm$0.02 & 0.81$\pm$0.01 \\
                     & TRUST-TSE & \bfseries 64.31$\pm$1.71 & \bfseries 1.39$\pm$1.33 & \bfseries 18.87$\pm$0.49 & \bfseries 18.44$\pm$0.54 & \bfseries 2.44$\pm$0.03 & \bfseries 3.10$\pm$0.07 & \bfseries 3.08$\pm$0.07 & 0.68$\pm$0.01 & \bfseries 0.93$\pm$0.01 & \bfseries 0.94$\pm$0.01 \\
\midrule
\multirow{3}{*}{DTU}  & NeuroHeed & 50.11$\pm$0.51 & -0.25$\pm$0.13 & 5.01$\pm$0.10 & 5.03$\pm$0.11 & 1.88$\pm$0.04 & 2.10$\pm$0.06 & 2.11$\pm$0.05 & 0.72$\pm$0.01 & 0.81$\pm$0.01 & 0.81$\pm$0.01 \\
                      & M3ANet    & 49.88$\pm$0.71 & -0.50$\pm$0.18 & 4.93$\pm$0.17 & 4.96$\pm$0.18 & 1.95$\pm$0.03 & 2.22$\pm$0.05 & 2.23$\pm$0.06 & \bfseries 0.73$\pm$0.01 & 0.83$\pm$0.01 & 0.82$\pm$0.01 \\
                     & TRUST-TSE & \bfseries 67.16$\pm$8.96 & \bfseries 3.03$\pm$4.46 & \bfseries 18.89$\pm$0.82 & \bfseries 18.09$\pm$0.73 & \bfseries 2.55$\pm$0.19 & \bfseries 3.12$\pm$0.07 & \bfseries 3.07$\pm$0.08 & 0.70$\pm$0.05 & \bfseries 0.89$\pm$0.01 & \bfseries 0.89$\pm$0.01 \\

\bottomrule
\end{tabular}
\label{tab:unseen_sub}
\end{table*}

\subsection{Linear probing on TRUST-TSE embeddings}
We follow the linear probing protocol in Section 3.3 to assess trial-discriminative information in TRUST-TSE EEG embeddings. Concretely, we freeze the EEG encoder and train a linear 8-way classifier to predict the KUL trial index, evaluating on temporally disjoint held-out segments from the same trials.

\begin{figure}[b]
  \centering
  \includegraphics[width=0.75\linewidth]{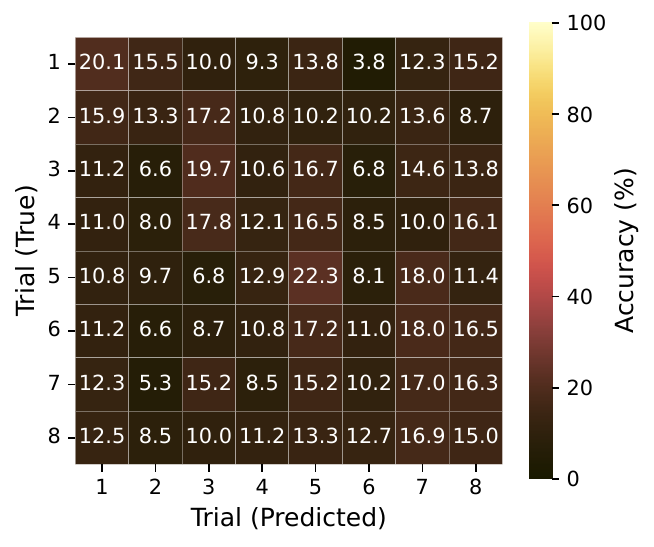}
  \caption{Confusion matrix for KUL trial-index prediction from frozen TRUST-TSE EEG embeddings via linear probing.}
  \label{fig:ours_probing}
\end{figure}

Figure~\ref{fig:ours_probing} shows that linear trial-index probing on frozen TRUST-TSE EEG embeddings achieves 16.3\% accuracy averaged across the eight trials, close to the 12.5\% chance level. Moreover, for four of the eight trials, the most frequent prediction is an incorrect trial, indicating that trial identity is not linearly decodable from TRUST-TSE embeddings.

\subsection{Confidence weight analysis}

\begin{figure}[!h]
  \centering
  \includegraphics[width=0.75\linewidth]{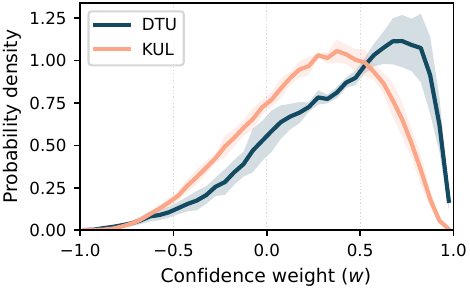}
  \caption{Distribution of confidence weights on the training sets of the KUL and DTU datasets. Solid lines show the mean and shaded regions indicate standard deviation across 8 runs.}
  \label{fig:cwhist}
\end{figure}

Figure~\ref{fig:cwhist} visualizes the empirical distribution of confidence weights $w$ on the training sets of DTU and KUL, computed segment-wise using the pretrained and frozen EEG encoders. Since $w=\tanh(\kappa\Delta)\in[-1,1]$ is a bounded monotone transform of the similarity margin $\Delta$, the plot provides a compact view of how frequently the EEG guidance is (i) strongly supportive ($w\rightarrow 1$), (ii) weak or ambiguous ($w\approx 0$), or (iii) contradictory to the trial-level attended target ($w<0$) over training segments. The distributions differ between the two datasets, with KUL containing a larger proportion of negative-weight segments than DTU.

\begin{table}[!b]
  \centering
  \caption{Summary statistics of negative confidence weights on the cross-trial training sets (mean$\pm$std over 8 runs).}
  \footnotesize
  \label{tab:cw_prob}
  \setlength{\tabcolsep}{12pt}
  \begin{tabular}{lcc}
    \toprule
    & \textbf{KUL} & \textbf{DTU} \\
    \midrule
    $P(w<0)$ & $0.256 \pm 0.024$ & $0.168 \pm 0.042$ \\
    $P(w<-0.25)$ & $0.105 \pm 0.012$ & $0.068 \pm 0.020$ \\    
    $P(w<-0.5)$ & $0.026 \pm 0.004$ & $0.022 \pm 0.008$ \\
    $P(w<-0.75)$ & $0.002 \pm 0.001$ & $0.003 \pm 0.002$ \\
    \bottomrule
  \end{tabular}
\end{table}

For a quantitative summary, Table~\ref{tab:cw_prob} reports the proportions of negative confidence weights on the training sets. The fraction of negative-weight segments is higher on KUL than DTU ($P(w<0)=0.256$ vs.\ $0.168$), indicating more frequent segment-level disagreement between the frozen embedding guidance and the trial-level supervision in KUL. At the same time, strongly negative weights are rare in both datasets ($P(w<{-}0.5)=0.026$ and $0.022$), implying that high-confidence contradictions constitute only a small portion of training segments.

These distributions help interpret the ablation trends in Section~5.4.2. Policies that discard all negative-weight segments effectively remove a non-trivial fraction of KUL training data (about one quarter on average), while the corresponding reduction is smaller for DTU. This observation is also qualitatively consistent with the discussion in Section~5.4.2 that maintaining sustained attention may be more challenging in KUL than DTU, which could lead to a higher rate of attention fluctuation within a trial. Nonetheless, since instantaneous attention labels are unavailable, negative-weight segments should not be regarded as direct evidence of attention switches.

\subsection{Effect of freezing the EEG encoder}

To examine whether Stage~2 should update the pretrained EEG encoder, we compare three training schemes in Table \ref{tab:freezing}. \textit{Frozen} corresponds to TRUST-TSE: Stage 1 pretrains the EEG encoder $F_\theta$ (together with the audio encoder $H_\psi$) using the cross-modal contrastive loss $\mathcal{L}_\text{NCE}$ and Stage~2 trains the EEG-conditioned extractor $G_\phi$ with $F_\theta$ fixed under the confidence-weighted SI-SDR objective. \textit{Fine-tuned} keeps the same Stage~1 pretraining but allows $F_\theta$ to be updated by the Stage~2 SI-SDR loss jointly with $G_\phi$. \textit{End-to-end (joint)} optimizes $\mathcal{L}_\text{NCE}$ and the SI-SDR loss simultaneously throughout training, updating $F_\theta$, $H_\psi$, and $G_\phi$ together.

Across both datasets, \textit{Frozen} yields the most reliable cross-trial target selection accuracy, achieving 62.27$\pm$2.51\% on KUL and 70.40$\pm$1.42\% on DTU, while \textit{Fine-tuned} drops substantially (54.71$\pm$6.93\% and 59.65$\pm$4.26\%, respectively). \textit{End-to-end (joint)} also underperforms $\textit{Frozen}$ and is notably less stable on KUL (55.22$\pm$14.23\%), suggesting brittle optimization when the EEG encoder must simultaneously satisfy EEG--speech alignment and separation objectives. Overall, these results indicate that propagating the Stage~2 SI-SDR gradients into the EEG encoder can erode the shortcut-resistant structure learned in Stage~1, whereas freezing preserves the pretrained alignment and improves both robustness to unseen trials and training stability. 

\begin{table}[h]
  \caption{Stage~2 accuracy (\%) for different training schemes.}
  \label{tab:freezing}
  \centering
  \footnotesize
  {\setlength{\tabcolsep}{9pt}
  \begin{tabular}{cccc}
  \toprule
    & \shortstack{\textbf{Frozen}\\~} & \shortstack{\textbf{Fine-tuned}\\~} &
    \shortstack{\textbf{End-to-end}\\\textbf{(joint)}} \\
  \midrule
  KUL
    & \bfseries 62.27 $\pm$ 2.51 & 54.71 $\pm$ 6.93   & 55.22 $\pm$ 14.23 \\
  DTU
    & \bfseries 70.40 $\pm$ 1.42 & 59.65 $\pm$ 4.26 & 64.29 $\pm$ 3.01 \\
  \bottomrule
  \end{tabular}}
\end{table}

\subsection{Hyperparameter tuning for envelope baselines}

In Section 5.4.3, we observed that deep envelope-decoding baselines (VLAAI and ADTNet) did not outperform mTRF and TRUST-TSE (envelope) under our cross-trial evaluation. To ensure a fair comparison, we performed dedicated hyperparameter tuning for mTRF, VLAAI, and ADTNet using the same cross-trial 4-fold setting as in the main experiments. For each candidate configuration, hyperparameters were selected by maximizing the validation Stage~1 selection accuracy, computed as the mean across the four folds with a fixed random seed. The final test accuracies reported in Table~7 were then obtained by re-evaluating the selected configuration with 4 folds $\times$ 4 random seeds.

For VLAAI, we tuned the number of blocks and the output context (OC) layer kernel size, as the original work highlights both the stacked block design and the OC layer as key architectural components affecting decoding performance, with the default configuration using 4 blocks and an OC layer kernel size of 32 samples. For ADTNet, we tuned the embedding dimension and the number of transformer blocks, as the original paper evaluates these hyperparameters and reports performance comparisons across different block counts, while keeping other settings fixed. For mTRF, we tuned the ridge regularization coefficient $\lambda$ via a log-spaced sweep.

\begin{figure}[!t]
  \centering

  \begin{subfigure}[t]{\linewidth}
    \centering
    \includegraphics[width=\linewidth]{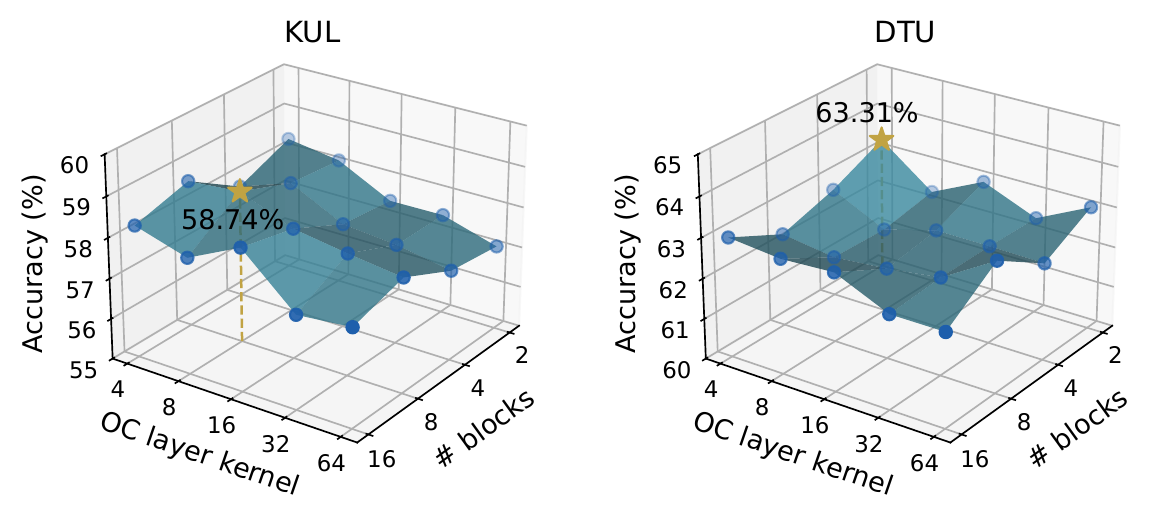}
    \caption{VLAAI}
    \label{fig:VLAAI_hp}
  \end{subfigure}

  \vspace{0.5em}

  \begin{subfigure}[t]{\linewidth}
    \centering
    \includegraphics[width=\linewidth]{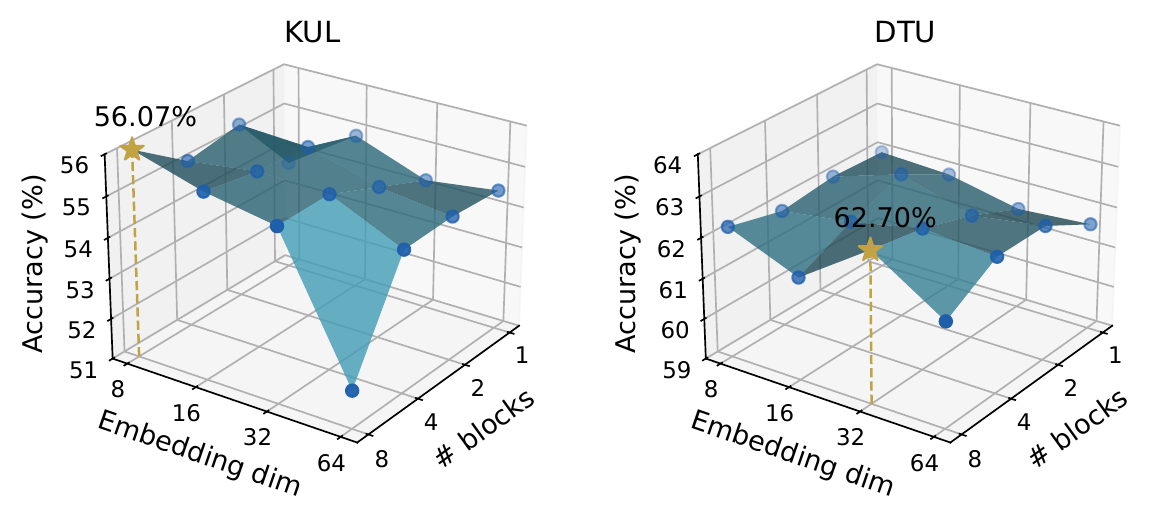}
    \caption{ADTNet}
    \label{fig:ADTNet_hp}
  \end{subfigure}

  \vspace{0.5em}

  \begin{subfigure}[t]{\linewidth}
    \centering
    \includegraphics[width=\linewidth]{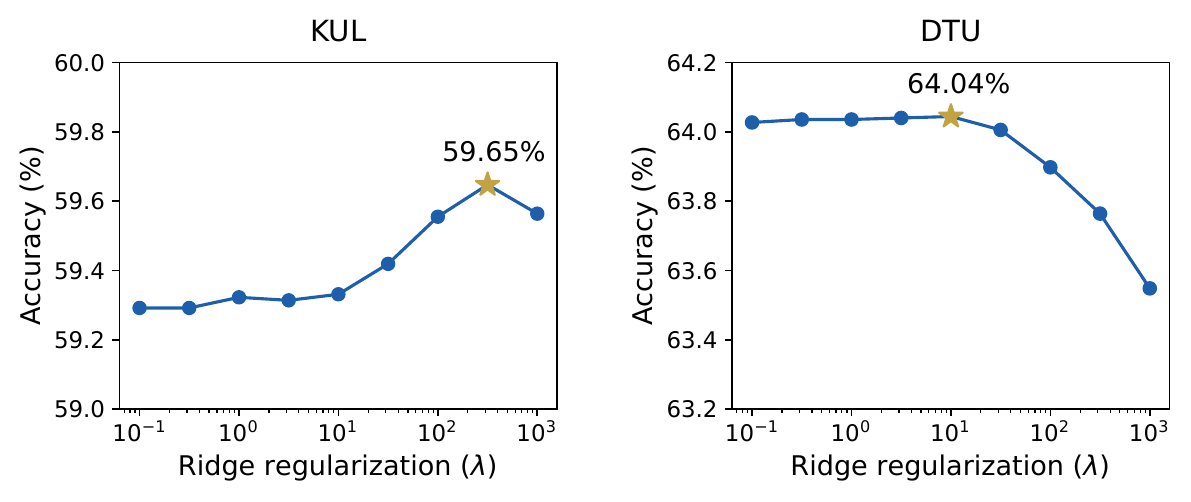}
    \caption{mTRF}
    \label{fig:mTRF_hp}
  \end{subfigure}

  \caption{Validation Stage~1 selection accuracy on KUL (left) and DTU (right), averaged over the 4 cross-trial folds with a fixed random seed. The star marker indicates the best validation configuration used in the main experiments.}
  \label{fig:env_hp}
\end{figure}

Figure~\ref{fig:env_hp} summarizes the hyperparameter tuning results for the envelope-decoding baselines. Notably, increasing model capacity does not monotonically improve validation accuracy for the deep baselines, which is consistent with the generalization challenges under cross-trial evaluation. Overall, these tuning results support that the baseline numbers reported in Table~7 are obtained under carefully selected hyperparameters rather than from arbitrary choices. To further control for architectural differences, we also include TRUST-TSE (envelope), an architecture-matched variant of TRUST-TSE (latent) that shares the same EEG encoder architecture while optimizing an envelope reconstruction objective.

